\documentclass[aps,pra,twocolumn,showpacs,superscriptaddress,groupedaddress]{revtex4}
\usepackage{amsmath}
\usepackage{graphicx}  
\usepackage{dcolumn}   
\usepackage{bm}        
\usepackage{amssymb}   

\usepackage{comment}

\newcommand\e{\varepsilon}

\newcommand{\lr}\longrightarrow
\newcommand{\ra}\rightarrow

\begin{document}

\title{Measures of quantum computing speedup}

\author{Anargyros Papageorgiou}
\email{ap@cs.columbia.edu} \affiliation{Department of Computer Science, Columbia University,
New York, NY 10027, USA.}

\author{Joseph F. Traub}
\email{traub@cs.columbia.edu}   \affiliation{Department of Computer Science, Columbia University,
New York, NY 10027, USA.}

\begin{abstract} 
We introduce the concept of strong quantum speedup. We prove that
approximating the ground state energy of an instance of the
time-independent Schr\"odinger equation, with $d$ degrees of freedom, $d$ large,
enjoys strong exponential quantum speedup. It can be easily solved on a quantum computer.
Some researchers in discrete complexity theory believe that quantum computation is not effective for eigenvalue problems. 
One of our goals
in this paper is to explain this dissonance.
\end{abstract}


\maketitle

\section{Quantum speedup criteria for computational science}

How much faster are quantum computers than classical computers for important problems? 
We shall show that the answers depend critically on how quantum speedup is measured.

We begin with two criteria for quantum speedup. \newline
{\bf Criterion 1.} For a given problem, the quantum speedup, $S_1$, is the ratio between the cost of the best classical algorithm
{\em known}\ and the cost of a quantum algorithm. That is
$$S_1=\frac{\rm cost\ of\ the\ best\ classical\ algorithm\ known}{\rm cost\ of\ a\ quantum\ algorithm}.$$
The cost of an algorithm is the amount of resources it uses, such as the number of arithmetic operations, or
elementary quantum operations, the number of oracle calls, etc.

Our second criterion uses the concept of computational complexity
which, for brevity, we will call complexity. By complexity we mean the minimal cost of solving a problem. \newline
{\bf Criterion 2.} For a given problem, the strong quantum speedup, $S_2$, is the ratio between classical complexity and 
quantum complexity. That is,
$$S_2=\frac{\rm classical\ complexity}{\rm quantum\ complexity}.$$

We are particularly interested in finding problems where $S_1$ or $S_2$ are exponential functions of a parameter.
We then refer to exponential quantum speedup and strong exponential quantum speedup, respectively.
The crucial difference between the two criteria is that if a problem satisfies Criterion 1 someone may invent a 
better classical algorithm and therefore decrease $S_1$ but if a problem satisfies Criterion 2 then $S_2$ cannot be 
decreased.

We apply these criteria to a well-known problem. Interest in quantum computing received a major boost in 1994 
\cite{SH94} when Peter Shor showed that factoring a large integer, $N$, can be done with cost $poly log$ in $N$.
The best classical algorithm known, the general number field sieve, has super-polynomial 
cost 
$$O(2^{c(\log N)^{1/3} (\log\log N)^{2/3}}).$$
Thus integer factorization has super-polynomial quantum speedup.
However, it is not known to have strong super-polynomial quantum
speedup. Although deciding whether an integer is prime can be done with 
polynomial cost \cite{AKS04}, it is still an open problem to determine
whether there exists an efficient algorithm for computing a
non-trivial factor of a composite  integer. This
is a problem in the complexity class FNP, which  is a functional analogue of the class
NP for decision problems; see \cite{P94}. At the moment it
is only conjectured that there is no polynomial factoring classical
algorithm; it is possible that
some day a polynomial-cost
factorization algorithm will be found.

Often scientists prove exponential quantum speedup for a problem but not strong exponential quantum speedup
because the classical complexity is not known. Examples include 
\cite{AL99,Aspuru-Guzik:2005yq,JLP11,JP03,LWa99,Wang:2008,WBA11}.
Establishing exponential quantum speedups marks an important advance in understanding the power of a
quantum computer.

We discuss three problems which enjoy strong quantum speedups. We emphasize that we do not claim separations of the 
complexity hierarchy in the sense $P\ne NP$.
Separation questions lead to open problems which are believed to be very hard. Our
complexity estimates and speedups are obtained using specific kinds of oracle calls.

The first problem is integration of functions of $d$ variables, which have uniformly bounded first partial derivatives. This is a
typical continuous problem in computational science. Since such problems have to be discretized we have only partial
information in the computer about the mathematical integrand. We obtain this partial information
using  oracles, for example, function evaluations. 

The cost of an algorithm is equal to the number of oracle calls plus the number of
operations the algorithm uses to combine the values returned by the oracle calls to obtain the approximate answer. 
In this sense, the oracle calls can be viewed as the input to the algorithm and the combinatory cost 
is a function of the number of oracle calls. 

The problem complexity is the minimal cost of any algorithm for solving the problem to within error $\e$. Let $m(\e,d)$
be the number of oracle calls for solving the problem. This is called the information complexity. Clearly the
information complexity is a lower bound on the complexity. This lower bound
does not require knowledge of the combinatory cost of an algorithm. For our integration problem $m(\e,d)$
is of order $\e^{-d}$. Thus the problem suffers the curse of dimensionality on a classical computer.

On a quantum computer the cost of the algorithm is the number of queries and other operations  required to solve the problem
to within $\e$ with probability greater than $1/2$. As before, the problem complexity is the minimal cost of any 
algorithm. The query complexity, i.e., the minimum number of quantum queries, provides a lower bound 
on the problem complexity. For our example of multivariate integration the quantum complexity is $o(\e^{-2})$
\cite{No01,He03b}. The curse of dimensionality is vanquished by a quantum computer. This problem enjoys  
strong exponential quantum speedup.

The second problem is a discrete example which also uses oracle calls. The problem is to compute the mean of a Boolean function, $f$,
of $n$ variables to within error $\e$. The minimal number of evaluations of $f$ , i.e., the number of oracle calls, is 
$2^{n-1}(1-\e)$ in the worst case.
The amplitude estimation algorithm of Brassard et al. \cite{BHMT02} can be used to compute the mean with
$O(\e^{-1})$ queries, for all $n\gg \log \e^{-1}$, plus a polynomial in $n$ and $\log \e^{-1}$ number of quantum operations 
that are required for the remaining parts of the algorithm excluding the queries. 
Thus this problem also enjoys strong exponential quantum speedup.

Finally, we mention that Simon's problem \cite{Si97} is solved using quantum queries and enjoys 
a strong exponential quantum speedup.

In the next section we will analyze the quantum speedup of approximating the ground state energy of the time-independent 
Schr\"odinger equation under certain conditions on the potential $V$. We will argue that this problem also enjoys strong exponential quantum
speedup. Some researchers in discrete complexity theory believe that quantum computation is not effective for eigenvalue problems. 
We will try to resolve this dissonance.

\section{Computing the ground state energy}

Consider the time-independent Schr\"odinger equation 
\begin{eqnarray*}
-\Delta u(x) + V(x) u(x) = \lambda u(x)  &&x\in I_d:=(0,1)^d \\
u(x) = 0 && x \in\partial I_d.
\end{eqnarray*}
Thus the equation is defined on the unit cube in $d$ dimensions with a Dirichlet boundary condition.
As usual, 
$$\Delta= \sum_{j=1}^d \frac{\partial^2}{\partial x_j^2}$$ 
denotes the Laplacian operator and $V\ge 0$ is the potential. Assume $V$
is uniformly bounded by unity, i.e., $\|V\|_\infty\le 1$, that it is continuous and has continuous first partial derivatives 
$D_jV:=\partial V/\partial x_j$, 
$j=1,\dots,d$, which satisfy $\|D_jV\|_\infty\le 1$. The problem is to approximate the ground state energy (i.e, the smallest eigenvalue) 
with relative error $\e$, using function evaluations of $V$.

Using perturbation arguments it was shown in
\cite{Pap07} that the ground state energy approximation is at least as hard as 
multivariate integration for $V$ satisfying the conditions above. 
Using lower bounds for multivariate integration \cite{NW08}
we conclude 
that on a classical computer with a worst case guarantee the complexity is at least proportional to 
\begin{equation}
\label{eq:detClb}
(cd\e)^{-d} \quad {\rm\ as\ } d\e\to 0,
\end{equation}
where $c>1$ is a constant. In fact, this lower bound follows from the number of evaluations
of $V$ (oracle calls) that are necessary for accuracy $\e$.
It is assumed that $V$ is known only at the evaluation points.
The complexity is an exponential function of $d$.
Moreover, discretizing the partial differential operator on a regular grid with mesh size $\e$
leads to a matrix eigenvalue problem, where the matrix has size $\e^{-d}\times \e^{-d}$. The evaluations of $V$ appear
in the diagonal entries. Inverse iteration can be used to approximate the minimum eigenvalue with
a number of steps proportional to $d\log\e^{-1}$.

The cost on a quantum computer is of order 
$$d\e^{-(3+\delta)} {\rm\ for\ any\ } \delta>0,$$ 
and the number of qubits is proportional to
$$d\log\e^{-1}.$$ 
The algorithm uses queries returning evaluations of $V$ truncated to $O(\log \e^{-1})$ bits. 
Its cost includes the number of queries and 
quantum operations.
See \cite{PPTZ13} for details. 
We remark that for a number of potentials $V$, such as the ones given by polynomials or trigonometric functions satisfying  
our assumptions, the queries can be implemented by quantum circuits of size
polynomial in $d\log \e^{-1}$.
The cost of the quantum algorithm provides an upper bound for the complexity. A lower bound of order 
$(d\e)^{-1/2}$ for the quantum complexity
follows by considering $V$ to be the sum of $d$ univariate functions. From the upper and lower quantum complexity bounds
one obtains a range for the anticipated speedup. 
Recall that the criterion for strong quantum speedup is 
$$S_2=\frac{\rm classical\ complexity}{\rm quantum\ complexity}.$$
Hence,
\begin{eqnarray*}
S_2 &=& O\left(  \frac{(cd\e)^{-d}}{(d\e)^{-1/2} } \right) \\
S_2 &=& \Omega\left(  \frac{(cd\e)^{-d}}{d
\e^{-(3+\delta) } } \right). {\rm\ as\ } d\e\to 0.
\end{eqnarray*}
Thus quantum computation enjoys strong exponential quantum  speedup, in $d$, over the worst case deterministic classical computation.

Another way of describing this is to say that quantum computation vanquishes the curse of (exponential) dimensionality. The reason for the
word in parentheses is that Richard Bellman coined the phrase curse of dimensionality informally in the preface of \cite{Be57}
in the study of 
dynamic programming for the solution of optimization problems. He noted that as the number of state variables increases solving the 
problem gets harder. This was an empirical result since it was before computational complexity theory was created. He did not specify
how the difficulty of the problem depended on the number of state variables which is why we inserted the word exponential.

\subsection{Comparison with QMA-complete problems}

Several researchers have written us (private communications) that strong exponential quantum speedup for the ground state energy problem cannot 
be true because it would imply that we have established complexity class separations.
We do not claim such separation results.

To illustrate this issue we will describe the assumptions appropriate for the continuous problems of computational science
comparing them to those for discrete problems.

We begin with the computational complexity of continuous problems. This is studied in the field of information-based complexity. 
There are numerous monographs and surveys describing the foundations and results. See, for example, 
\cite{TWo80,No88,TWW88,Pl96,TWe98,NW08,NSTW09,NW10,NW12}.

Typically, these problems are solved numerically and therefore approximately to within error $\e$. In applications the user chooses
the appropriate $\e$. Another key parameter is $d$ which denotes the degrees of freedom or dimension of the problem. The size of the
input, the algorithm cost and the problem complexity depend on $\e$ and $d$.

We specialize to the ground state energy problem. The continuous problem has structure which can be exploited by the quantum algorithm. 
The Hamiltonian is $-\tfrac12 \Delta + V$, where $\Delta$ is the Laplacian and $V$ is the potential. 
In the previous section we specified the conditions on $V$.

After discretization 
the resulting matrix is the sum of two matrices.
One involves the discretized Laplacian $\Delta_\e$ while the other is a diagonal matrix $V_\e$ which contains function
evaluations of the potential $V$. All the properties of $\Delta$ and $\Delta_\e$ are well known as well as the relationship
between them. This includes the eigenvalues, their separation, their distribution,
the eigenvectors etc. \cite{Cou89,For04,Tit58}. 
In particular, for small $\e$ the smallest eigenvalue of $-\tfrac 12 \Delta + V$ is close to that of $-\tfrac 12 \Delta_\e + V_\e$. 
The discretization has preserved the problem structure.
Most importantly, 
since the norm of $V_\e$ is
small the eigenvalues of $-\tfrac 12 \Delta_\e+ V_\e$ are close to those 
of $-\tfrac 12 \Delta_\e$. 

Turning to discrete problems, 
the complexity class
QMA is the quantum analogue of NP. It is also called BQNP and the name QMA was given 
by Watrous \cite{Wa00}. 
A decision problem about the ground state energy of local Hamiltonians is a QMA-complete problem \cite{Kit02,Kit06,AN02}
and this has been a very important and influential result in discrete complexity theory.
Hence, it is believed it is very hard to solve.
A number of papers state that approximating the ground state energy is a QMA-hard problem; see, e.g., \cite{JGL10,WMN10}.
A similar conclusion holds for a density functional theory approach for eigenvalue approximation \cite{SV09}.

There are important differences between our ground state problem and the local Hamiltonian QMA-complete problem.
The deterministic worst case complexity lower bound of equation (\ref{eq:detClb}) has been obtained using an oracle
and cannot provide a lower bound for the complexity of the QMA-complete local Hamiltonian problem whose
input is provided explicitly. For the same reason it cannot imply a separation in the complexity hierarchy.
Furthermore, it is unlikely that  
the quantum algorithm of \cite{PPTZ13} could be used to solve the local Hamiltonian problem. 
This algorithm was designed specifically to approximate the smallest eigenvalue of $-\tfrac 12 \Delta_\e+V_\e$. 
Using it for the local Hamiltonian
problem would require one to convert efficiently a sum of  
polynomially many local Hamiltonians to the sum of two matrices,
where we know everything about the first and the second is a diagonal matrix with relatively small norm. 
Observe that in the case of the local Hamiltonian problem the input is a polynomial number of local 
Hamiltonians all satisfying the same properties \cite[Def. 2.3]{Kit06}
without any apparent useful distinction between them. 
While it is easy to deal with each Hamiltonian individually, 
dealing with their sum is a hard problem.

Moreover, the qubit complexity of our ground state problem is proportional to $d\log\e^{-1}$ and so is
the number of qubits used by the quantum algorithm. The number of qubits for the QMA complete local Hamiltonian problem is 
denoted by $n$
in \cite[Def. 2.3]{Kit06} and the sum of the local Hamiltonians is a $2^n\times 2^n$ matrix. 
Note that in the former case the number of qubits is derived from the parameters $\e$ and $d$, while in the latter case $n$ is the 
parameter.
For the QMA-complete problem it is not known if there exists a quantum algorithm solving it with cost polynomial in $n$.

Relating the cost of the quantum algorithm in \cite{PPTZ13} to the number of its qubits highlights further differences between the cost requirements
of discrete complexity theory and the continuous problems of physical sciences and engineering.
We have shown that the ground state energy can be approximated with cost polynomial in $d$ and $\e^{-1}$ which
makes it \lq\lq easy\rq\rq.
Yet from the point of view of discrete complexity theory such problems are considered hard. 
We believe discrete complexity theory sets too high a bar.
It requires that the cost of the quantum algorithm is a polynomial in the number of qubits (or, equivalently, polylogarithmic in the matrix size)
for the algorithm to be considered efficient 
(recall that in \cite[Def. 2.3]{Kit06} $n$ is the number of qubits upon which each Hamiltonian acts, 
the Hamiltonian is a $2^n\times 2^n$ matrix, and the
input size is polynomial in $n$.) See also \cite{BCC06}.

Here is the crux of why the bar is too high for the continuous problems of computational science. 
With the exception of problems such as the solution of nonlinear equations and convex optimization, 
scientific problems cannot be solved with cost polynomial in $\log\e^{-1}$ \cite{NW08}.
One of the simplest continuous scientific problems is approximating a univariate integral with the promise that the integrand has a 
uniformly bounded first derivative.
The classical complexity of this integration problem is $\Theta(\e^{-1})$ and the quantum complexity is $\Theta(\e^{-1/2})$;
see \cite{HNo02,No01} for more details. As we stated earlier the lower bound for the quantum complexity of the
ground state problem is proportional to $(d\e)^{-1/2}$.
Note that in computational science $\e$ is often not very small. A typical value is $\e=10^{-8}$. 
Thus complexity of order $\e^{-1}$ or $\e^{-1/2}$ is not expensive.
Note that the complexity of univariate integration is exponentially greater than $\log\e^{-1}$.

We have been asked by researchers in discrete complexity theory what if $\e$ is constant. As
stated above $\e$ and $d$ are parameters and one studies the complexity for all their values.
Secondly, this possibility is excluded for our ground state energy problem 
since then the problem is trivial. 
Indeed, unless $\e < 1/d$ the problem 
can be solved with constant cost by the algorithm that does not make any evaluations at all and 
returns the value $d\pi^2+1/2$, since then the error is bounded by $1/d$.

\section{Discussion}

If one looks at the approximation of the ground state energy of the $d$-dimensional Schr\"odinger equation through the prism of 
discrete complexity theory one would 
conclude that this problem is hard to solve on a quantum computer. 
We propose that the speedup measures $S_1$ or $S_2$ are more relevant for the power
of quantum computing for problems in computational science.

We suggest that analogous results might be found if one were to investigate other continuous problems.
That is they are tractable on a quantum computer from the point of view of computational science but intractable 
from the view of discrete complexity theory.

\section*{Acknowledgements}
This work has been supported in part by NSF/DMS.
We thank M. Yannakakis for his recommendation to highlight the differences in the models of computation.

\bibliography{pra-qspeedup}

\end{document}